\documentclass{pasa}%
\usepackage{rotating}
\usepackage[T1]{fontenc}
\usepackage{amssymb}
\usepackage{amstext}
\usepackage{float}
\usepackage{graphicx}
\usepackage{caption}
\usepackage{subcaption}
\usepackage{color}
\usepackage{array}
\usepackage{changepage}
\usepackage{longtable}
\usepackage{amsmath} 
\usepackage{nth}

\usepackage{aas_macros}
\usepackage{hyperref} 
\hypersetup{colorlinks,citecolor=blue,linkcolor=blue,urlcolor=blue}

\usepackage[authoryear]{natbib}
\bibpunct{(}{)}{;}{a}{}{,}
\DeclareRobustCommand{\ion}[2]{%
\relax\ifmmode
\ifx\testbx\f@series
{\mathbf{#1\,\mathsc{#2}}}\else
{\mathrm{#1\,\mathsc{#2}}}\fi
\else\textup{#1\,{\mdseries\textsc{#2}}}%
\fi}

\newcommand{\Ha}{\rm H$\alpha$}

\title[Performance of a novel polymer imaging bundle]{Performance of a novel PMMA polymer imaging bundle for field acquisition and wavefront sensing \vspace{-0.4cm}}
\author[Richards et al.]{{\large S.~N.~Richards$^{1,2,3,4}$\thanks{E-mail: \href{samuel@physics.usyd.edu.au}{samuel@physics.usyd.edu.au}},
S.~Leon-Saval$^{1,4}$, 
M.~Goodwin$^{2}$
J.~Zheng$^{2}$,
J.~S.~Lawrence$^{2}$,
J.~J.~Bryant$^{1,2,3}$,
J.~Bland-Hawthorn$^{1,4}$,
B.~Norris$^{1,4}$,
N.~Cvetojevic$^{1,2,4,5,6}$ \and 
A.~Argyros$^{6}$}\vspace{0.1cm}\\ 
\affil{$^{1}$Sydney Institute for Astronomy, School of Physics, University of Sydney, NSW 2006, Australia}%
\affil{$^{2}$Australian Astronomical Observatory, PO Box 915, North Ryde, NSW 1670, Australia}
\affil{$^{3}$CAASTRO: ARC Centre of Excellence for All-sky Astrophysics}
\affil{$^{4}$Sydney Astrophotonic Instrumentation Laboratory (SAIL), School of Physics, University of Sydney, Sydney, NSW 2006, Australia}
\affil{$^{5}$CUDOS: ARC Centre of Excellence, Centre for Ultrahigh bandwidth Devices for Optical Systems}
\affil{$^{6}$Institute of Photonics and Optical Science (IPOS), School of Physics, University of Sydney, Sydney, NSW 2006, Australia}}
\jid{PASA}
\doi{10.1017/pas.\the\year.xxx}
\jyear{\the\year}

\begin{document}%
\begin{abstract}
Imaging bundles provide a convenient way to translate a spatially coherent image, yet conventional imaging bundles made from silica fibre optics typically remain expensive with large losses due to poor filling factors ($\sim40$\%). We present the characterisation of a novel polymer imaging bundle made from poly(methyl methacrylate) (PMMA) that is considerably cheaper and a better alternative to silica imaging bundles over short distances ($\sim1$\,m; from the middle to the edge of a telescope's focal plane). The large increase in filling factor ($92$\% for the polymer imaging bundle) outweighs the large increase in optical attenuation from using PMMA ($1$\,dB/m) instead of silica ($10^{-3}$\,dB/m). We present and discuss current and possible future multi-object applications of the polymer imaging bundle in the context of astronomical instrumentation including: field acquisition, guiding, wavefront sensing, narrow-band imaging, aperture masking, and speckle imaging. The use of PMMA limits its use in low light applications (e.g. imaging of galaxies), however it is possible to fabricate polymer imaging bundles from a range of polymers that are better suited to the desired science.
\end{abstract}
\begin{keywords}
instrumentation: miscellaneous -- telescopes -- instrumentation: adaptive optics -- instrumentation: high angular resolution
\end{keywords}
\maketitle%
\section{INTRODUCTION}
\label{sec:intro}

Over the course of the past few decades, the act of bundling optical fibres together to translate an image in a spatially coherent manner has been optimised to increase the spatial resolution, yet retaining efficiency and flexibility. These devices are known as imaging fibre bundles (or coherent fibre bundles), and are widely used for remote sensing with applications such as biomedical endoscopy. Most fibre bundles are made using silica optical fibres, however the process used to make a $\sim1$\,mm diameter bundle that remains flexible sacrifices efficiency due to the fibre filling factor. This is because for most of the length, each fibre needs to be loose, otherwise a $1$\,mm bundle would be a solid rod with near zero flexibility. The {\sc SCHOTT} Leached Image Bundles\footnote{\href{http://www.schott.com/lightingimaging/english/medical/medical-products/transmitting-images\_leached-image-bundle.html}{http://www.schott.com/lightingimaging/english/medical/ \\ medical-products/transmitting-images\_leached-image-bundle.html}} are a good example of this, where the inter-core cladding material is etched down over the length of the fibre, leaving only the ends bundled together. The delicate process of making a flexible silica fibre bundle increases the cost to $\sim\rm US\$1000$~per~metre. High filling factor silica fibre bundles do exist \citep[e.g.][]{2011OExpr..19.2649B, 2014MNRAS.438..869B}, though their large individual core diameters ($\sim100\,{\rm \mu m}$) results in a coarse spatial resolution, and with the bundle only being fixed at one end, they lack the ability to preserve a spatially coherent image and therefore are not classed as imaging bundles.

The use of polymer as an alternative to silica in fibre imaging bundles presents an attractive solution to many of the issues faced by using silica (e.g. the ability to obtain a high filling factor whilst retaining flexibility). One of the most common polymer fibre materials is poly(methyl methacrylate) (PMMA), however the use of PMMA (and most polymers) over silica comes at the cost of attenuation (on the order of $1$\,dB/m for PMMA versus $10^{-3}$\,dB/m for silica). This increase in attenuation limits the applications of polymer fibre to cases where fibre lengths on order of a metre are required (e.g. endoscopy, localised remote sensing). \linebreak Albeit, longer lengths of these polymer imaging bundles can be used in applications where the source is of high enough power to overcome the attenuation loss.

An area where fibre optics continue to increase in their application is astronomy, where they have become the dominant method of translating the light of thousands of galaxies and stars from the telescope focal plane to nearby spectrographs. Most focal planes of current large ($4$-$10$\,m) telescopes are $\lesssim1$\,m in diameter, but there are soon to be focal planes of $1$-$2$\,m diameter upon the arrival of Extremely Large Telescopes (ELTs). These large focal planes require precise field acquisition and guiding, for which coherent imaging bundles are well suited. In this application, at one end, multiple imaging bundles are deployed over the focal plane targeting guide stars, and at the other end, all of the imaging bundles are brought together with one camera situated away from the focal plane imaging their end faces. This method has been adopted by the Sloan Digital Sky Survey multi-object spectrograph \citep[SDSS;][]{2013AJ....146...32S} and the Sydney-AAO Multi-object Integral field spectrograph \citep[SAMI;][]{2015MNRAS.447.2857B}. SDSS use silica imaging bundles from Sumitomo\footnote{Sumitomo Electric Lightwave Corp.,\\ \href{http://www.sumitomoelectric.com}{http://www.sumitomoelectric.com}}, and SAMI use polymer imaging bundles from ESKA\footnote{ESKA$^{\rm TM}$; Mitsubishi Corp., \href{http://fiberopticpof.com}{http://fiberopticpof.com}}. 

\begin{figure}
\centering
\includegraphics[width=0.8615\linewidth]{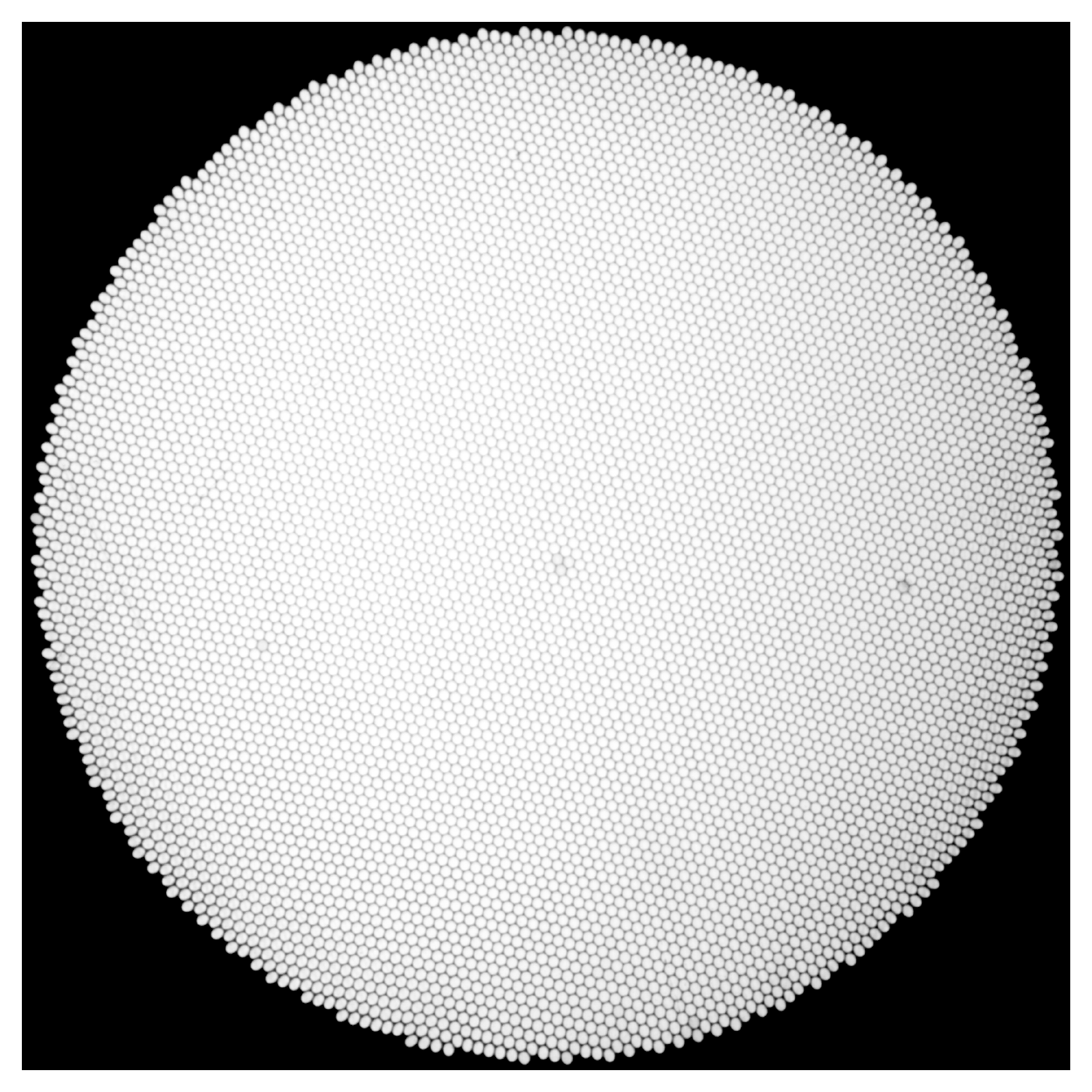}
\includegraphics[width=0.8615\linewidth]{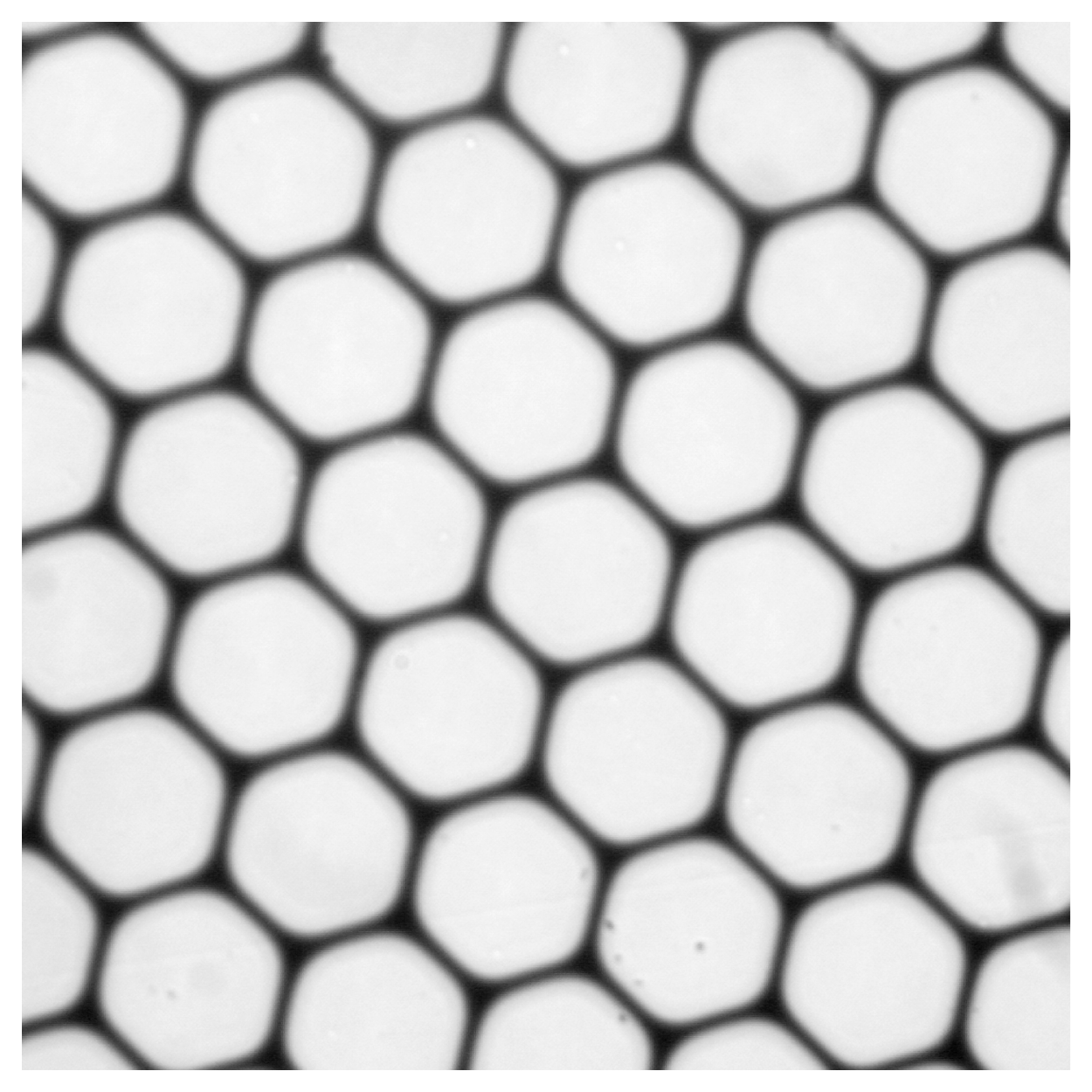}
\includegraphics[width=0.8615\linewidth]{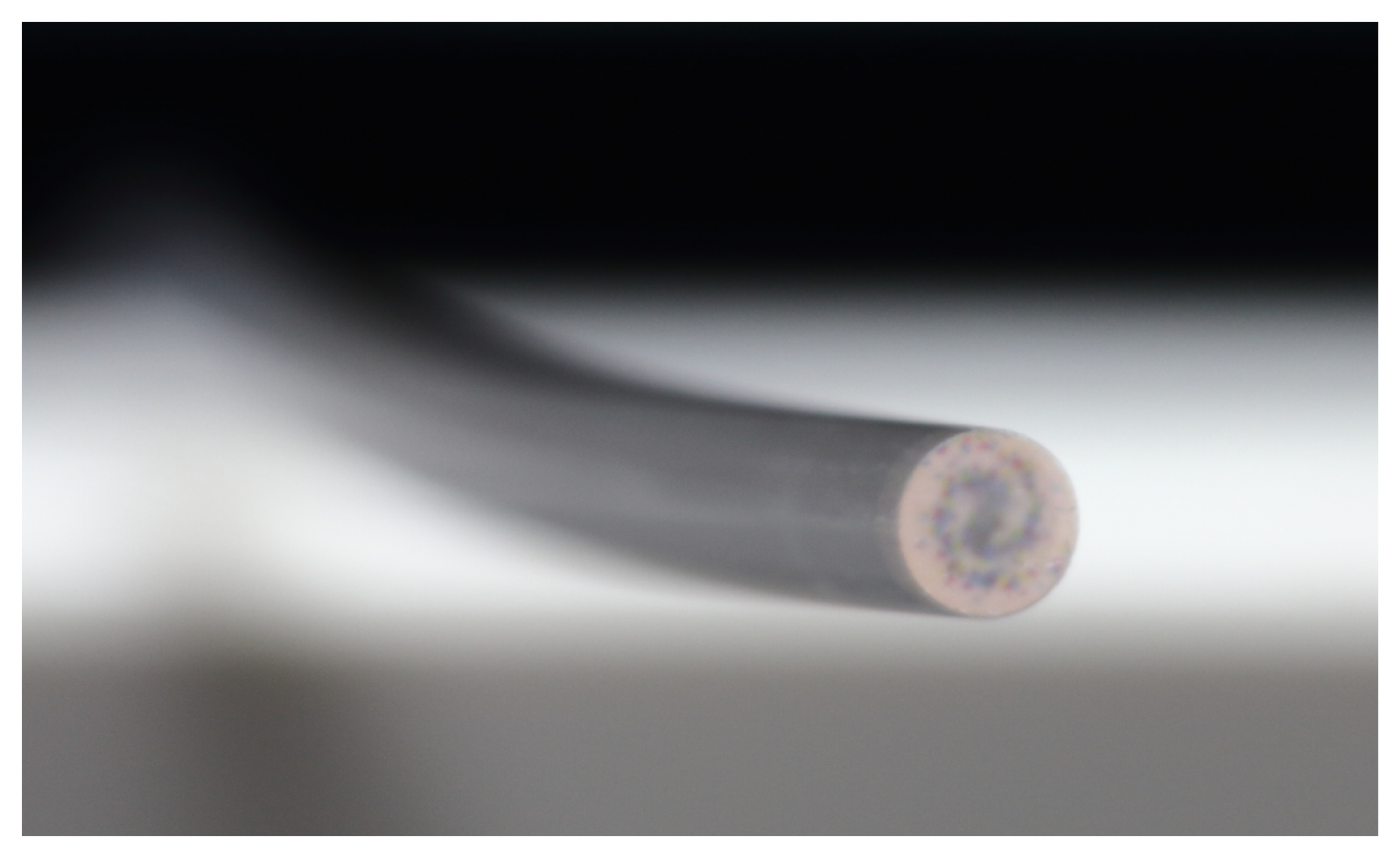}
\caption{(top) Back-illuminated microscope image of a coherent polymer bundle showing the entire face of the bundle. (middle) a magnified image of (top) showing the hexagonal structure of the cores. There are $7095$ cores hexagonally packed, with each core being $16\,{\rm \mu m}$ (full-diagonal) resulting in an overall diameter of $1.5$\,mm. Dust in the microscope imaging system gives the appearance of damaged cores. (bottom) a projected \Ha\ map of a spiral galaxy by butt-coupling the input end of the bundle to a $1.5$\,mm grayscale printout of the galaxy.}
\label{fig:PIB}
\end{figure}

Basic characterisation of the Sumitomo silica image bundles has been performed by \citet{2006SPIE.6273E..3UH}, who analysed the spatial fidelity, cross talk and scattering, and concluded that more investigation into imaging bundles is needed to assess their suitability in astronomical applications. The Sumitomo bundle has $2\,{\rm \mu m}$ cores spaced by $3\,{\rm \mu m}$ in a semi-regular hexagonal pattern, resulting in a filling factor of $\sim40\%$ \citep{2008OSA.47.25.4560U}. This immediate substantial loss from the filling factor is detrimental in astronomical applications of these spatially coherent imaging bundles, where every photon counts. In addition to the filling factor, core sizes of $2\,{\rm \mu m}$ are very close to being purely single-moded at optical wavelengths resulting in an even greater loss due to very poor modal coupling \citep{2014OSA.15.1443H}. In the case of the Sumitomo bundle, the outer diameter is only $800\,{\rm \mu m}$, meaning that even without separating the individual cores along the length of the fibre, the bundle retains some flexibility.

More recently, {\sc SCHOTT} have developed a Wound Fibre Bundle\footnote{SCHOTT Wound Fibre Bundle, \href{http://www.schott.com/lightingimaging/english/defenseproducts/wound.html}{http://www.schott.com/\\lightingimaging/english/defenseproducts/wound.html}} for the Defense sector, that creates a spatially coherent array of smaller $6\times6$ fibre arrays with each fibre being $10\,{\rm \mu m}$ in diameter. Its efficiency over the wavelength range $400-1500$~nm is typically $40$-$50$\%, including attenuation, fill-factor and Fresnel surface reflection losses. Detailed characterisation on its optical performance has yet to be carried out. 

Our interest then lies with the ESKA polymer imaging bundles (PMMA), which have a larger filling fraction than silica imaging bundles, yet retain spatial resolution and flexibility. In Section \ref{sec:char} we present the results of laboratory characterisations of two different sizes of polymer imaging bundles. In Section \ref{sec:app} we explore their various applications in the context of astronomical instrumentation, presenting recent on-sky demonstrations, and concluding our work in Section \ref{sec:conc}.


\vspace{0cm}

\section{Optical Performance} \label{sec:char}

To better understand the range of possible applications polymer imaging bundles have in astronomical instrumentation, we performed laboratory characterisation on their filling factor, flat-field response, spectral throughput, cross-talk and focal ratio degradation. We explored two sizes of ESKA polymer imaging bundles, with external diameters of $0.5$ and $1.5$\,mm. They are both made by the same method, with the $0.5$\,mm bundle being drawn down further in the process, resulting in the same number of cores, but smaller core diameters. Both bundles have $7095$ hexagonally-shaped cores, with each core diameter being $16\,{\rm \mu m}$ (full-diagonal) in the $1.5$\,mm bundle and $6\,{\rm \mu m}$ (full-diagonal) in the $0.5$\,mm bundle. In some of the tests we present the results of both bundle sizes, although we primarily tested the $1.5$\,mm bundle. A back-illuminated microscope image of the $1.5$\,mm bundle can be found in Figure \ref{fig:PIB}. In all tests, the end faces were machine polished\footnote{Krelltech Rev2 Micro Polisher} stepping down through $30,12,9,3,1,0.3\,{\rm \mu m}$ grit paper.


\vspace{0cm}

\subsection{Filling factor and flat-field} \label{sec:fill}

\begin{figure*}
  \begin{minipage}[t]{0.48\linewidth}
    \centering
    \includegraphics[width=\linewidth]{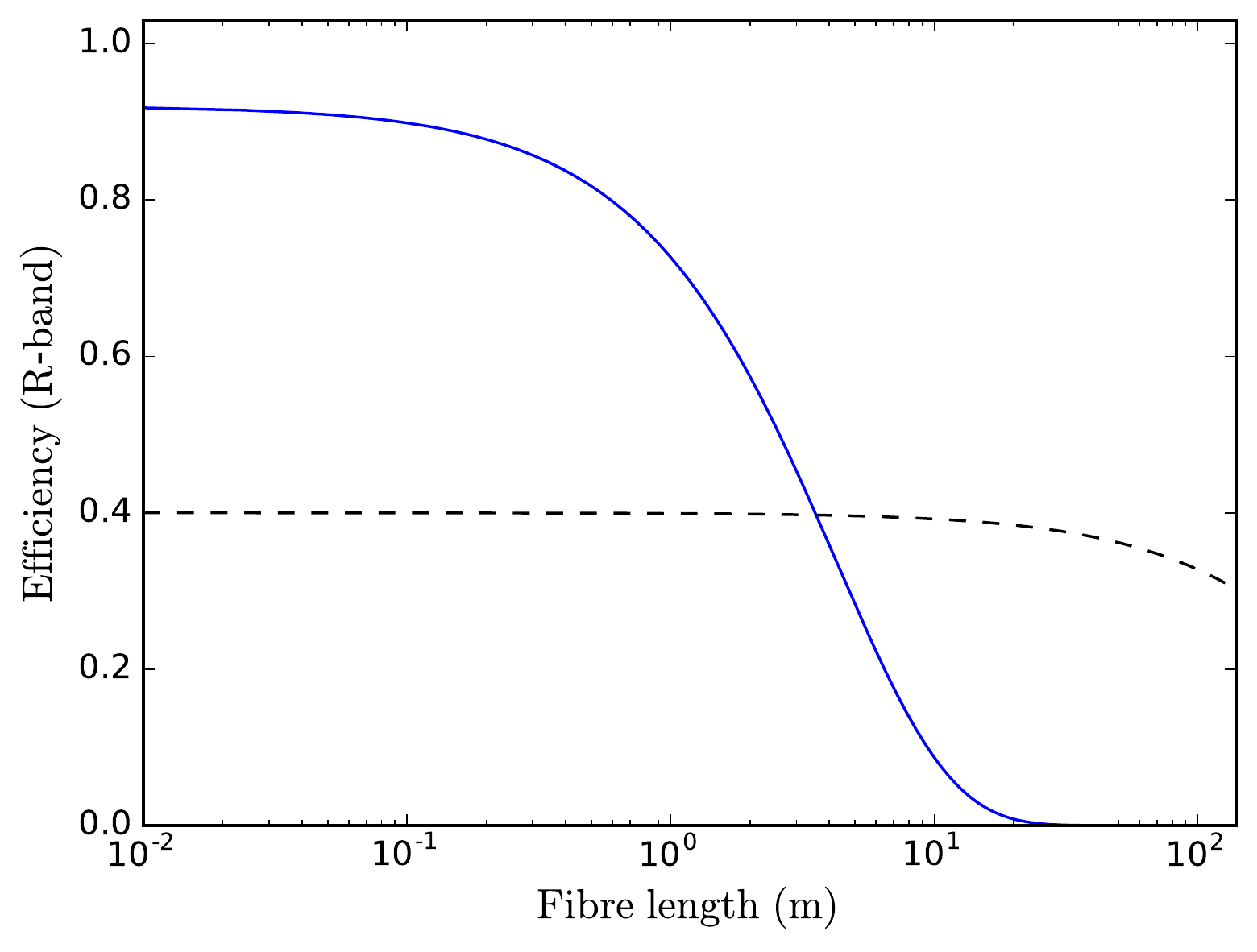} 
    \caption{Theoretical R-band efficiency as a function of fibre length for a $1.5$\,mm polymer imaging bundle (blue line) and a best silica imaging bundle (dashed line), where the efficiency includes the fibre attenuation ($1$\,dB/m and $0.01$\,dB/m respectively) and the fill factor ($92$\% and $40$\% respectively).} 
    \label{fig:PIB_vs_silica}
  \end{minipage} \hfill
  \begin{minipage}[t]{0.48\linewidth}
    \centering
    \includegraphics[width=\linewidth]{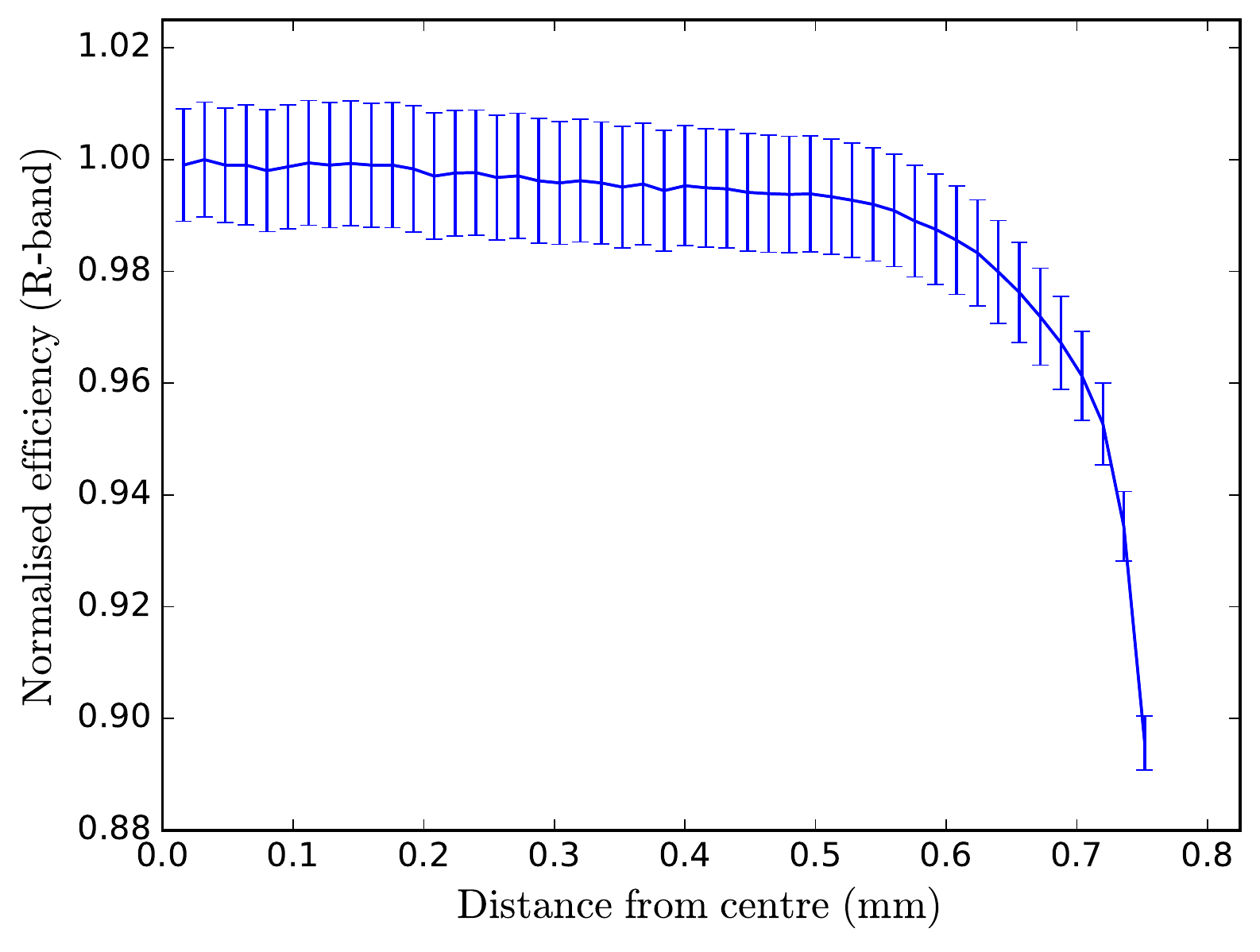} 
    \caption{A normalised radial profile of the $1.5$\,mm polymer imaging bundle. Each measurement is the mean intensity of pixels within annuli of widths $16\,{\rm \mu m}$ placed on an R-band back-illuminated flat-field image of the bundle. Error bars represent the standard deviation of each measurement.\vspace{0.2cm}}
    \label{fig:PIB_FF}
  \end{minipage}
  \begin{minipage}[t]{\linewidth}
    \centering
	\includegraphics[width=\linewidth]{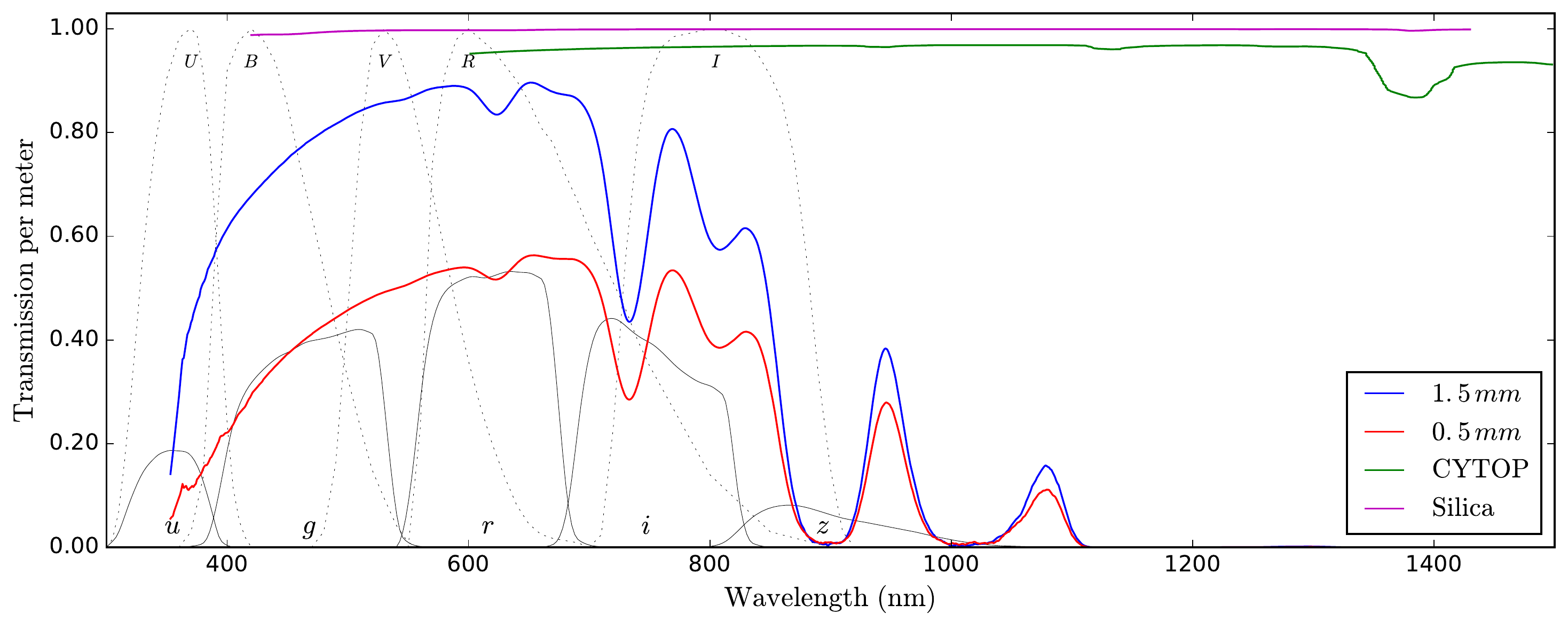} 
    \caption{Spectral throughput of the $0.5$ and $1.5$\,mm polymer imaging bundles (red and blue respectively). The green line is the transmission profile of a new polymer material, CYTOP, taken from \citet{2013ISRN.785162..22A}, and the magenta line is the manufacture's transmission of AFS200220T silica fibre. Bessel-Johnson (UBVRI; dotted black line) and SDSS (ugriz; solid black line) spectral filters are overlaid for reference (arbitrarily scaled). There was no transmission of wavelengths less than $\sim350$\,nm and greater than $\sim1100$\,nm for either bundle. \vspace{-0.2cm}} 
    \label{fig:PIB_transmission} 
  \end{minipage}  
\end{figure*}

The dominant source of losses in fibre bundles, in terms of total efficiency, is a poor filling factor, where the total area at the bundle end-face that can be sampled by the cores is less than the overall area. As the cores are what guides the incident light and maintain the spatial coherence (with the rest of the bundle consisting of the cladding), any light not coupled is at best lost, or at worst couples into the cladding that contributes to cross-talk. Due to the small core sizes found in imaging bundles, sufficient cladding to confine propagating modes results in larger separations between cores. If the space between cores becomes too small then the modes can easily leak into adjacent cores, spreading out the spatial information and increasing light losses. Creating a large refractive index difference between the core and the cladding allows the core-to-core separation to be smaller, which is possible to achieve when using polymer rather than silica (see Section \ref{sec:FRD}). By analysing a front-illuminated image of the $1.5$\,mm polymer imaging bundle, the filling factor was measured to be $92$\%. This is a factor of $2$-$3$ improvement over the best silica imaging bundles ($25-40$\%). This gain in filling factor is critical for astronomy, where missing spatial information can be detrimental to the science. When the filling factor of this polymer imaging bundle is compounded with the typical attenuation of PMMA ($1$\,dB/m), it is easy to see, from Figure \ref{fig:PIB_vs_silica}, that in short lengths ($<4$\,m) the polymer imaging bundles theoretically out-performs even the best silica imaging bundles ($0.01$\,dB/m and $40$\% filling factor). 

By evenly back-illuminating the bundle, using an OLED flat-screen source through a Bessel-Johnson R-band filter, the flat-field response profile of the $1.5$\,mm polymer imaging bundle was found by taking the mean of image pixel intensities within radially increasing annuli (see Figure \ref{fig:PIB_FF}). Each annulus has a width of $16\,{\rm \mu m}$ (one core), with the image initially dark subtracted and then subtracted by the maximum inter-core intensity to reduce the contamination of cladding light. The radial profile is observed to have a slight decrease of $\sim1$\% out to a radius of $\sim600\,{\rm \mu m}$ before decreasing rapidly to $\sim90$\% at the edge of the bundle. We conjecture that this type of radially decreasing profile is likely linked to the intrinsic scattering and crosstalk properties of the bundle studied in Section \ref{sec:FRD}. Linear propagation losses resulting in scattered light out of the central cores will result in an increased crosstalk between neighbouring cores, however the same scattering mechanism in the outer diameter cores will result in power loss out of the composite waveguide, i.e. bundle.  Variations in intensity within the annuli, in addition to the radial profile, are easily corrected for by flat-fielding in a data reduction process. The $1.5$\,mm polymer imaging bundle is therefore deemed acceptable for astronomical applications that require accurate flux information of a source. This is true even when the source is extended spatially.

\vspace{-0.4cm}
\subsection{Spectral transmission}

Although the use of polymer can increase the filling factor of imaging bundles dramatically (see Section \ref{sec:fill}), it has much larger optical attenuation compared to silica ($1$\,dB/m and $10^{-3}$\,dB/m respectively). To see how the attenuation varies as a function of wavelength, the spectral throughput of both polymer imaging bundles was measured via the cut-back method (see Figure \ref{fig:PIB_transmission}). A reference spectrum was obtained by coupling a white-light source into a $200\,{\rm \mu m}$ core-diameter silica patch cord and measured with an optical spectrum analyser (OSA). End-face polished ($0.3\,{\rm \mu m}$ grit) and connectorised lengths of the imaging bundles were then butt-coupled to the patch-cord and fed into the OSA. The lengths of imaging bundles were $10,5,2,1$\,m, re-polishing and re-connectorising after each cut. The median spectral transmission per~metre of each bundle was determined using the measurements of the multiple lengths normalised by the reference spectrum. The absorption bands seen in the spectral transmission curve (Figure \ref{fig:PIB_transmission}) are due to the molecular vibrations of the PMMA, becoming essentially opaque to light of wavelengths shorter than $\sim350$\,nm and longer than $\sim1100$\,nm. The aforementioned losses that arise from the single-mode nature of the cores in the $0.5$\,mm bundle are clearly evident in the scaling difference between the transmission of the two bundles, resulting in an additional loss of $\sim40$\%. Both transmission curves have been corrected for Fresnel end face losses. Figure \ref{fig:PIB_transmission} also shows the spectral transmissions for a new polymer, CYTOP\footnote{CYTOP, Asahi Glass Co:\\ \href{http://www.agc.com/kagaku/shinsei/cytop/en/}{http://www.agc.com/kagaku/shinsei/cytop/en/}} (Cyclic Transparent Optical Polymer), and a single silica fibre (AFS200220T\footnote{Anhydroguide Fiber Silica (AFS), Fiberguide Industries: \\ \href{http://www.fiberguide.com/product/optical-fibers/}{http://www.fiberguide.com/product/optical-fibers/}}).


\begin{figure*}
\centering
\centerline{\includegraphics[width=\linewidth]{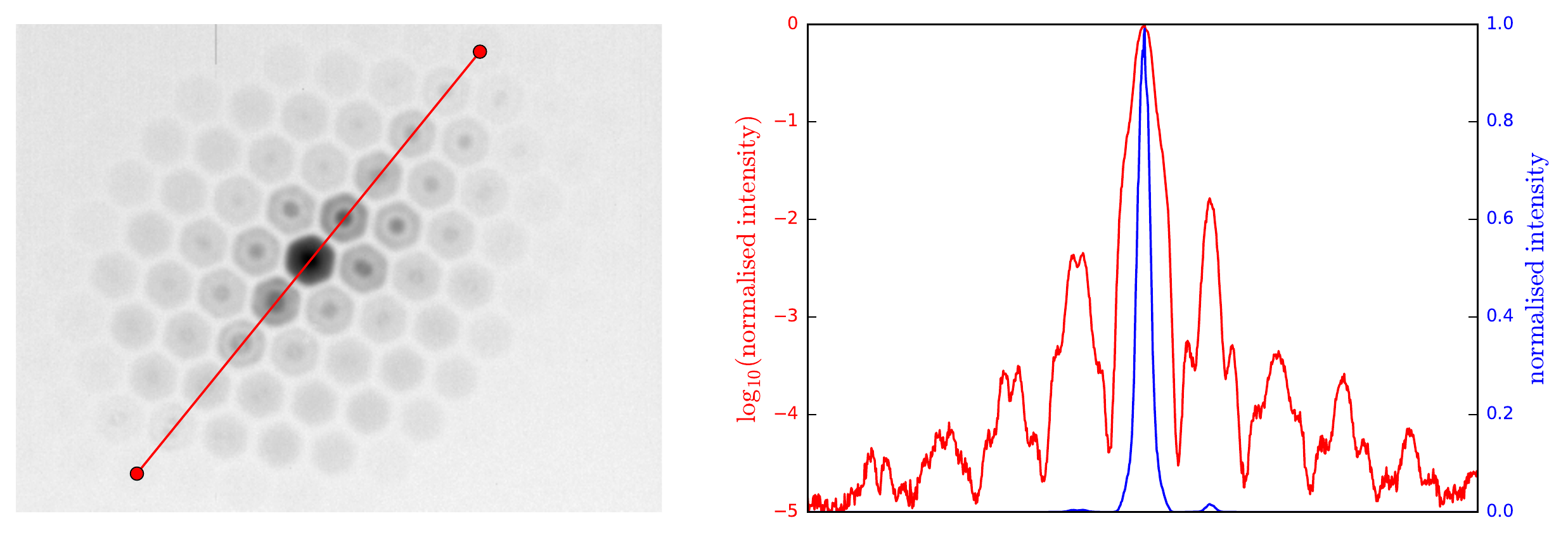}}
\caption{(left) A log scaled inverted image after butt-coupling and aligning a silica single-mode fibre to the centre of one core. (right) The intensity profile of the cut shown overlaid on (left) going from left to right. The red line is the log scaled intensity profile and the blue line is intensity on a linear scale. Both lines have been normalised to the maximum of the injected core. $3$\% of the injected light has been scattered to the surrounding cores. \vspace{0.0cm}}
\label{fig:PIB_scatter}
\end{figure*}

\subsection{Focal ratio degradation, cross-talk and scattering} \label{sec:FRD}

Due to the chemical composition and imperfections of optical fibres, conservation of the input light is difficult. This is even more so in standard polymers (e.g. PMMA); as demonstrated by the high attenuation levels when compared to silica, due to the higher scattering of the polymer media. Unlike when making optical fibres from silica, the ability to mix polymers together makes it possible to increase the numerical aperture of the fibre by creating a larger refractive index difference between the core and the cladding. This allows for a higher number of modes to propagate down the fibre and better constrains the mode field diameters. However, the chemical structures of PMMA and silica are very different, leading to a much larger scattering term for PMMA than silica. This increase in scattering also increases the chance of coupling between different modes. Light scattering can enhance the coupling from low order modes to higher order modes closer to the core mode cut-off that are weakly guiding and lightly to couple to cladding modes. When fibres are brought close together, as in the case of imaging bundles, the scattered light that leaves the core has a chance of coupling into its neighbouring cores. If the propagating modes are poorly confined due to a small difference in refractive index between the core and the cladding, the propagating light has a high chance of evanescently coupling to neighbouring cores. These coupling terms are known as cross-talk, and minimising this property is important for imaging bundles as to not spread out the spatial information of the input light source. The coupling of modes to higher order modes that were not coupled to at the input as well as the lading of modes to the cladding is known as focal ratio degradation (FRD), where the cone angle (numerical aperture; NA) of the input light becomes larger at the output of the fibre (or that the encircled energy of a given profile is less at the output than at the input light). 

\begin{figure*}
\begin{minipage}{\linewidth}
\centering
{\setlength{\fboxsep}{0pt}%
\setlength{\fboxrule}{0.5pt}%
\fbox{\includegraphics[width=0.325\linewidth]{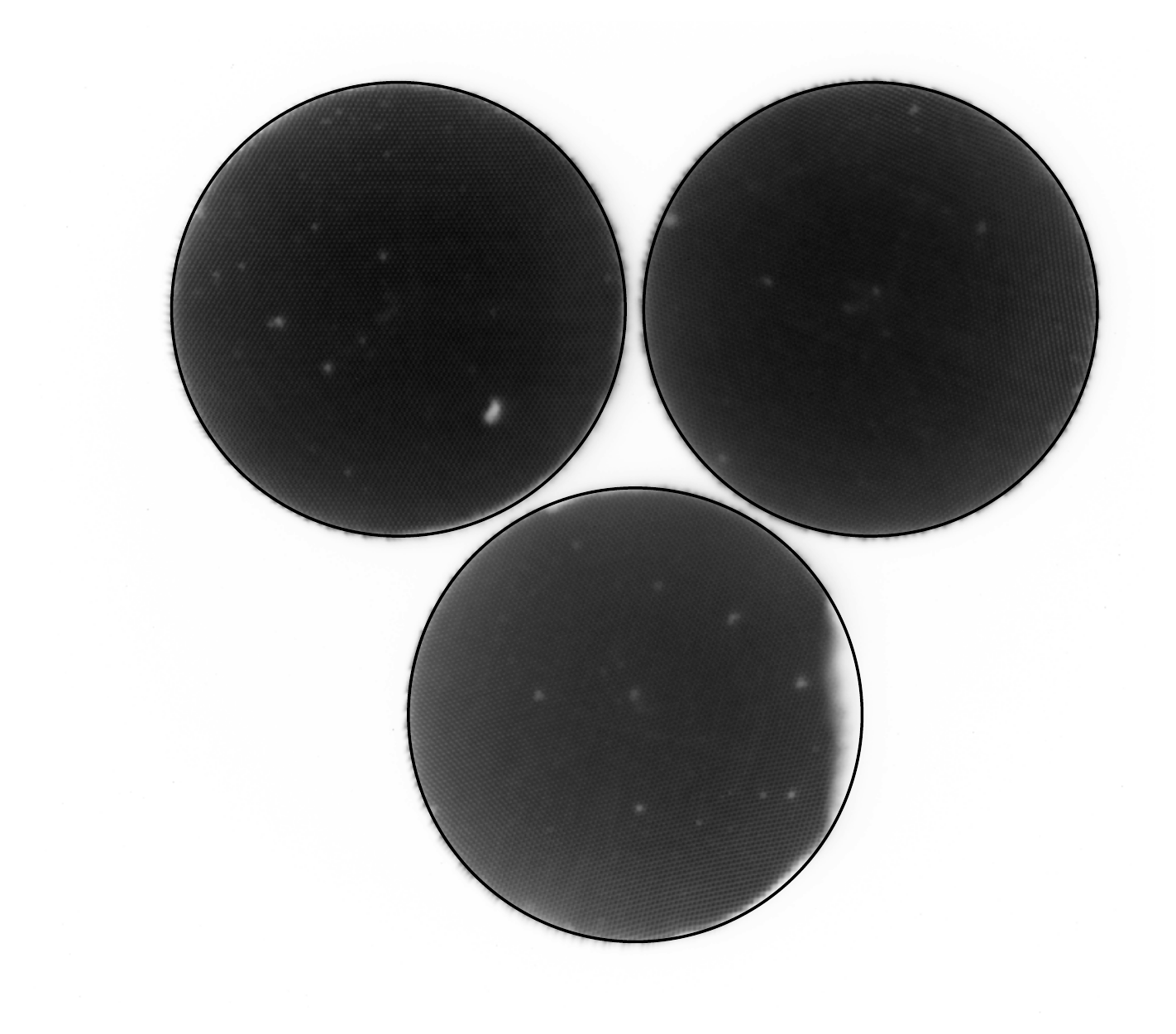}}} \hfill
{\setlength{\fboxsep}{0pt}%
\setlength{\fboxrule}{0.5pt}%
\fbox{\includegraphics[width=0.325\linewidth]{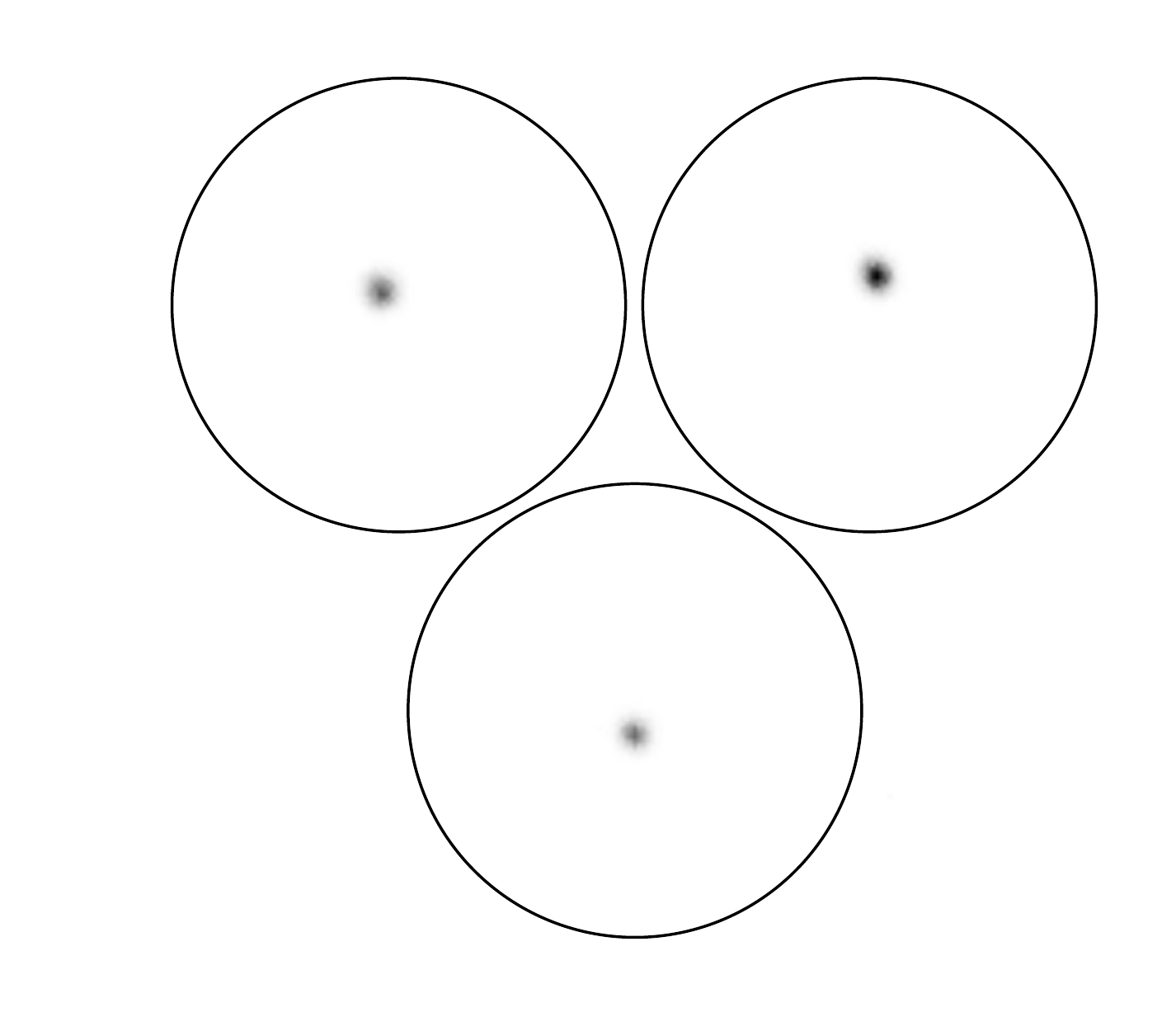}}} \hfill
{\setlength{\fboxsep}{0pt}%
\setlength{\fboxrule}{0.5pt}%
\fbox{\includegraphics[width=0.325\linewidth]{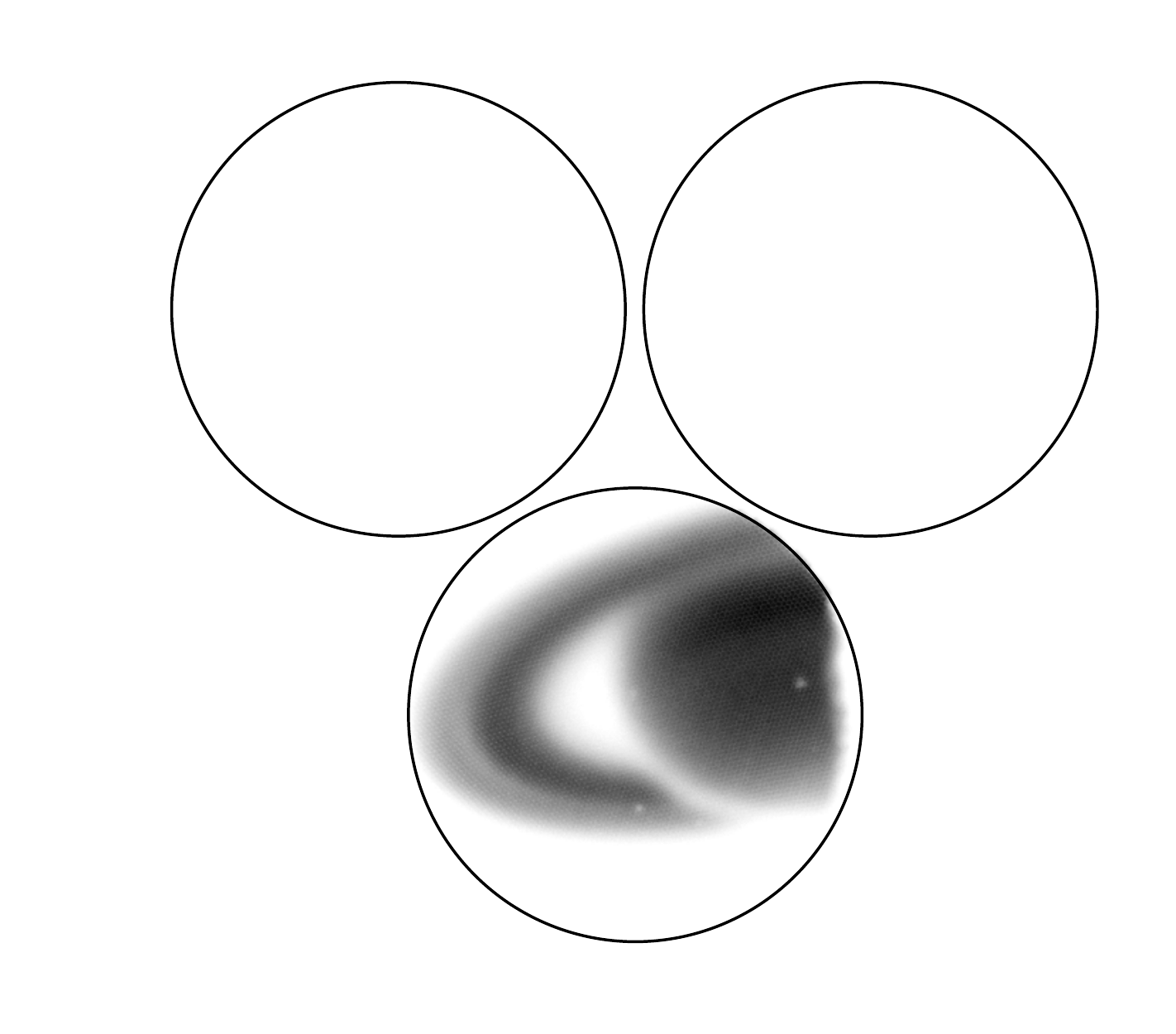}}}
\caption{Three $1.5$\,mm polymer imaging bundles used as guiding probes for SAMI. (left) sky-flats of the three bundles imaged through a $1.5$:$1$ re-imaging lens to a an Apogee Alta U6 CCD. Dust in the imaging system gives the appearance of damaged cores, however the lower bundle is damaged on its right side. (middle) three guide stars being probed by the polymer imaging bundles. Typical V-band magnitudes and exposures times for SAMI guide stars are \mbox{$13$\,-\,$15$} and $1$\,s respectively. (right) image of Saturn and its rings through one of the bundles ($\sim2$~arcsec seeing). Each imaging bundle has a $23$~arcsec field-of-view. The outline of each bundle is traced by a black circle. All three images are inverted. \vspace{0.4cm}}
\label{fig:PIB_guiding}
\end{minipage}
\end{figure*}

Knowledge of a fibre's FRD is important when designing an optical system that makes use of optical fibres. It dictates the power of the optical lenses used to image the near-field (face) of the fibre, whether as part of an imaging system or spectrograph. As the individual core diameters of the polymer imaging bundles are small ($6\,{\rm \mu m}$ and $16\,{\rm \mu m}$), it was not possible to measure the FRD via the standard method of injecting a focussed spot of light (of varying NAs) down a single core and measuring the NA of the output light \citep{2014MNRAS.438..869B}. Another technique to measured FRD is to inject collimated light at varying angles, but this too requires injection into a single core. Therefore, no FRD measurement was taken, but the large acceptance NA of the polymer imaging bundle would indicate that its FRD would be not too dissimilar to that of a standard silica fibre. As an absolute reference, the maximum NA of both polymer imaging bundles was measured by overfilling the bundle's input NA (in the R-band), and radially measuring the far-field intensity using a power meter with a $500\,{\rm \mu m}$ pinhole mask at a radial distance of $20$\,cm. This size of pinhole mask was chosen to be large enough for the power meter to sufficiently detect the incident light, yet small enough to provide discrete measurements at the various angles and shield the power meter from scattered light in the laboratory. The maximum NA (at $5$\% maximum intensity) of the $0.5$\,mm and $1.5$\,mm polymer imaging bundle is $0.46\pm0.03$ and $0.49\pm0.03$ respectively, with the errors representing the range of NAs found at intensity values either side of the $5$\% maximum intensity.  Therefore, due to the small overall diameter of these imaging bundles, designing an optical system to image the face of the bundle, or even $100$ bundles packed together, with $f/1$ optics (NA = $0.5$) is straightforward.

To measure the cross-talk in the $1.5$\,mm polymer imaging bundle, a silica single-mode fibre (NA = $0.1$, injected with R-band light) was butt-coupled and aligned to the centre of a single core, and the intensity of all of the surrounding cores was imaged (see Figure \ref{fig:PIB_scatter}). It was found that $2.9\pm0.2$\% of the injected light was scattered into the surrounding cores. This is the mean measurement and standard deviation of five different locations over the face of the imaging bundle (with the exception of the outermost cores).

Due to the high-NA, i.e. large mode confinement, of the polymer imaging bundles, it is unlikely that the modes' evanescent fields are significant to allow efficient coupling between cores, meaning that the primary driver for the cross-talk is more likely to be the large scattering term of the polymer's chemical structure. This is due to the larger molecular chains and clusters of the polymer in addition to the molecular absorptions of their vibrational states. Unlike a standard silica fibre when unspooled, the polymer imaging bundle retains a relaxed bend radius of $\sim15$\,cm. It is possible to straighten the fibre via thermal annealing, although doing so risks damaging the fibre (e.g. changing the polymer's chemical / physical structure). 


\section{Applications of polymer imaging bundles in astronomy} \label{sec:app}
In this section we present current and possible (but by no means exhaustive) applications of polymer imaging bundles in astronomical instrumentation that use lengths $<2$\,m. For this we consider the trade-offs of the increase in filling factor at the cost of an increase in attenuation. Fundamentally, polymer imaging bundles translate images, which in the context of astronomy is best suited to getting an image from the focal plane of a telescope to a nearby location where bulkier systems reside (cameras etc). The need of imaging bundles in astronomy increases as the diameter of the telescope's focal plane increases, more so in the era of the ELTs. 

\vspace{0.4cm}
\subsection{Field acquisition and guiding}

The most obvious application of imaging bundles is for telescope guiding as part of a multi-object instrument, or where the weight of a guide camera on the focal plane is an issue (e.g. small telescopes). It is possible to use pick-off mirrors on arms to do multi-object guiding, but if there are optical science fibres covering the field plate then it is difficult to avoid collisions. Imaging bundles permit a flexible solution to translating a spatially coherent image of a guide star from the focal plane to a nearby camera. An advantage of using polymer over silica for multi-object imaging bundles (other than the increase in filling factor) is that the end preparation of a polymer imaging bundle is much more customisable, as the product resembles a single fibre. Silica imaging bundles have to be carefully connectorised (normally by the manufacturer) to minimise stress or fractures, whereas the polymer imaging bundles can either be mounted in custom connectors, or just left bare whilst maintaining the ability to be polished. 

SAMI is a multi-object integral field spectrograph that deploys 13 silica fibre-bundle integral field units (hexabundles) over a $1$~degree field at the $3.9$\,m Anglo-Australian Telescope \citep{2015MNRAS.447.2857B}. For field acquisition and guiding it uses three of the $1.5$\,mm polymer imaging bundles ($1$\,m lengths and each with $23$~arcsec field-of-view) mounted in the same plug-plate connector design as the hexabundles. All of the imaging bundles feed a single guide camera located nearby to the side of the focal plane (see Figure \ref{fig:PIB_guiding}). Although only one of the guide stars is used for telescope guiding in SAMI, having multiple guide stars over a $1$~degree field allows for better understand of the wide-field atmospheric conditions, and the tip/tilt determination of the field plate.

The light-weight and flexible nature of the polymer imaging bundles makes them suited to many types of fibre positioning technologies. In addition to use in plug-plate instruments, the $1.5$\,mm polymer imaging bundle has also been proven to be compatible with the StarBugs positioning system \citep{2012SPIE.84501A.G,2014SPIE.9147E..10K}, and multiple $0.5$\,mm polymer imaging bundles will perform the guiding for the multi-object fibre spectrograph instrument, TAIPAN on the UK Schmidt Telescope \citep[][see Figure \ref{fig:PIB_starbug}]{2014SPIE.9147E..10K}.  Even though there is a loss in efficiency when using the $0.5$\,mm polymer imaging bundle instead of the $1.5$\,mm polymer imaging bundle, the smaller core sizes of the $0.5$\,mm polymer imaging bundle are better suited to the plate scale of the UKST ($0.44$ to $1.08$~arcsec~per~core respectively, with the latter being too coarse to accurately sample the guide star's point-spread-function). The $0.5$\,mm polymer imaging bundle is also being considered to replace the current guide bundles on the 2df instrument on the $3.9$\,m Anglo-Australian Telescope, which uses a pick-and-place positioning system with fibre retractors \citep{2002MNRAS.333..279L,2006SPIE.6269E..0GS}. Investigation into the stresses induced on the $0.5$\,mm polymer imaging bundle and its longevity when used in sprung retractors is underway.

As focal planes become increasingly larger, particularly in the era of ELTs where focal planes will be $1$-$2$\,m diameter, the need of fibre bundles to perform guiding also increases. This is where the adaptability of the polymer imaging bundles to different positioning technologies has an advantage, in addition to the large increase in filling factor. Typical plate-scales for the ELTs are a few arcsec/mm, meaning that the field-of-view of a $1.5$\,mm polymer imaging bundle would also be a few arcsec. Blind pointing of modern large telescopes is commonly $\sim1$~arcsec \citep{2014SPIE.9145E..1CB}, meaning that even without fore-optics, multiple $1.5$\,mm polymer imaging bundles are able to acquire and guide on a field. If an increased field-of-view per imaging bundle is desired, then the use of fore-optics (miniature lenses or fibre tapers) is straightforward. 


\vspace{0cm}
\subsection{Multi-object Wavefront Sensing}

The use of adaptive optics \citep[AO;][]{1953PASP...65..229B,1998aoat.book.....H} to correct for atmospheric turbulence is an essential part of the science cases proposed for ELTs, but is also a technology that is currently available on telescopes all over the world. Performing AO on a single on-axis target is now standard practice for $\geq8$\,m class telescopes, though recent studies into multi-object AO \citep[MOAO;][]{2002sdef.conf..139H,2010SPIE.7736E..0SR,2014A&A...569A..16V} show the power of maximising the scientific output of large aperture wide-field telescopes, such as the $25.4$\,m Giant Magellan Telescope \citep[GMT;][]{2014SPIE.9145E..1CB}. For MOAO to be realised on telescopes such as the GMT, many wavefront sensors would need to be deployed across the focal plane, localised to each science target. 

\begin{figure}
\begin{minipage}{\linewidth}
\centerline{\includegraphics[width=0.955\linewidth]{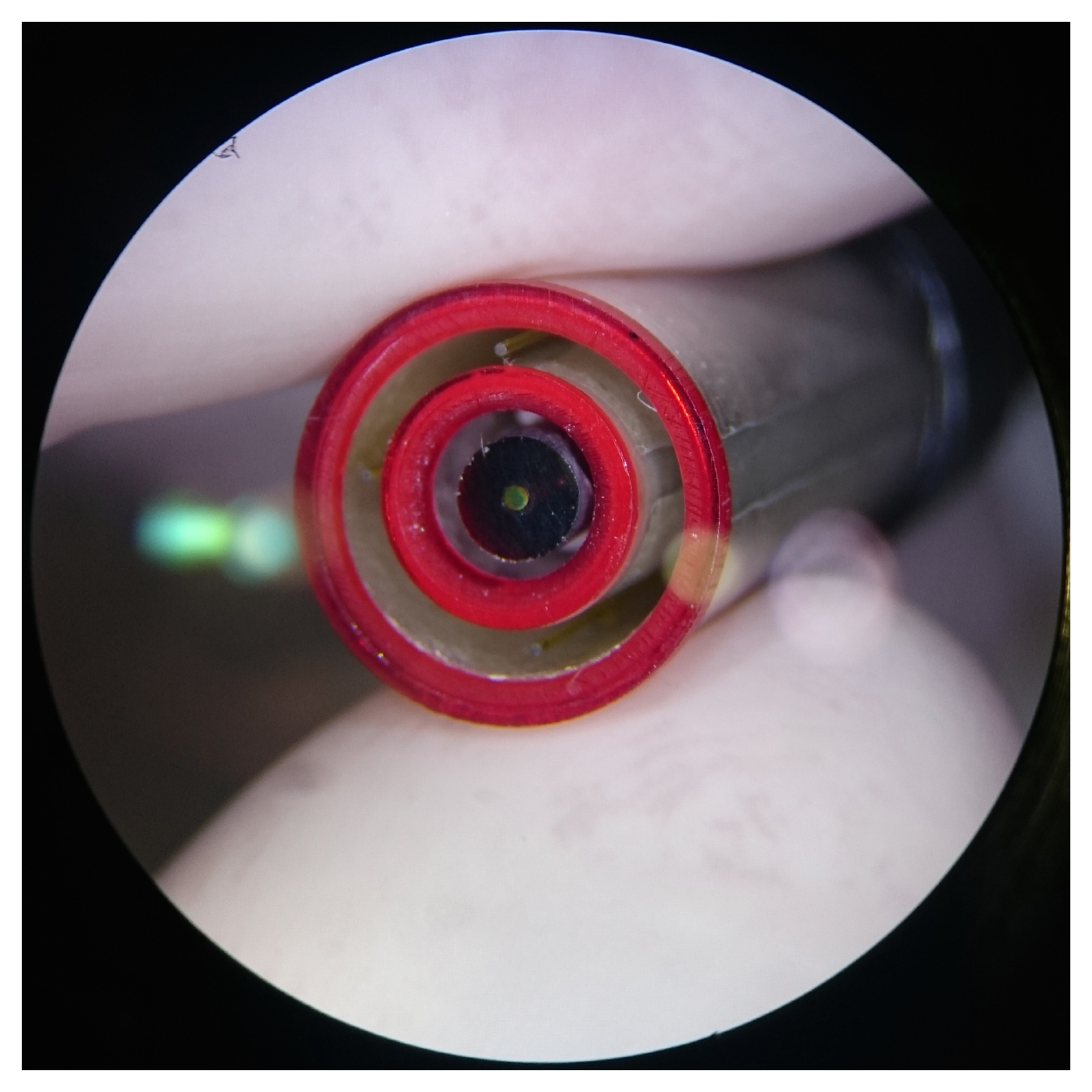}}
\caption{Microscope image of a 0.5mm diameter polymer imaging bundle mounted in a TAIPAN StarBug with a 3D-printed ferrule.}
\label{fig:PIB_starbug}
\end{minipage}
\begin{minipage}{\linewidth}
\centerline{\includegraphics[width=\linewidth]{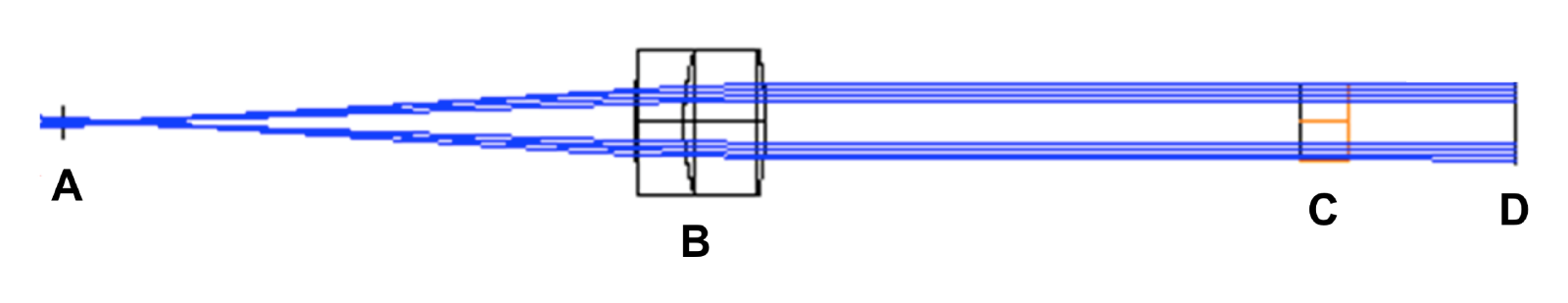}}
\centerline{\includegraphics[width=0.49\linewidth]{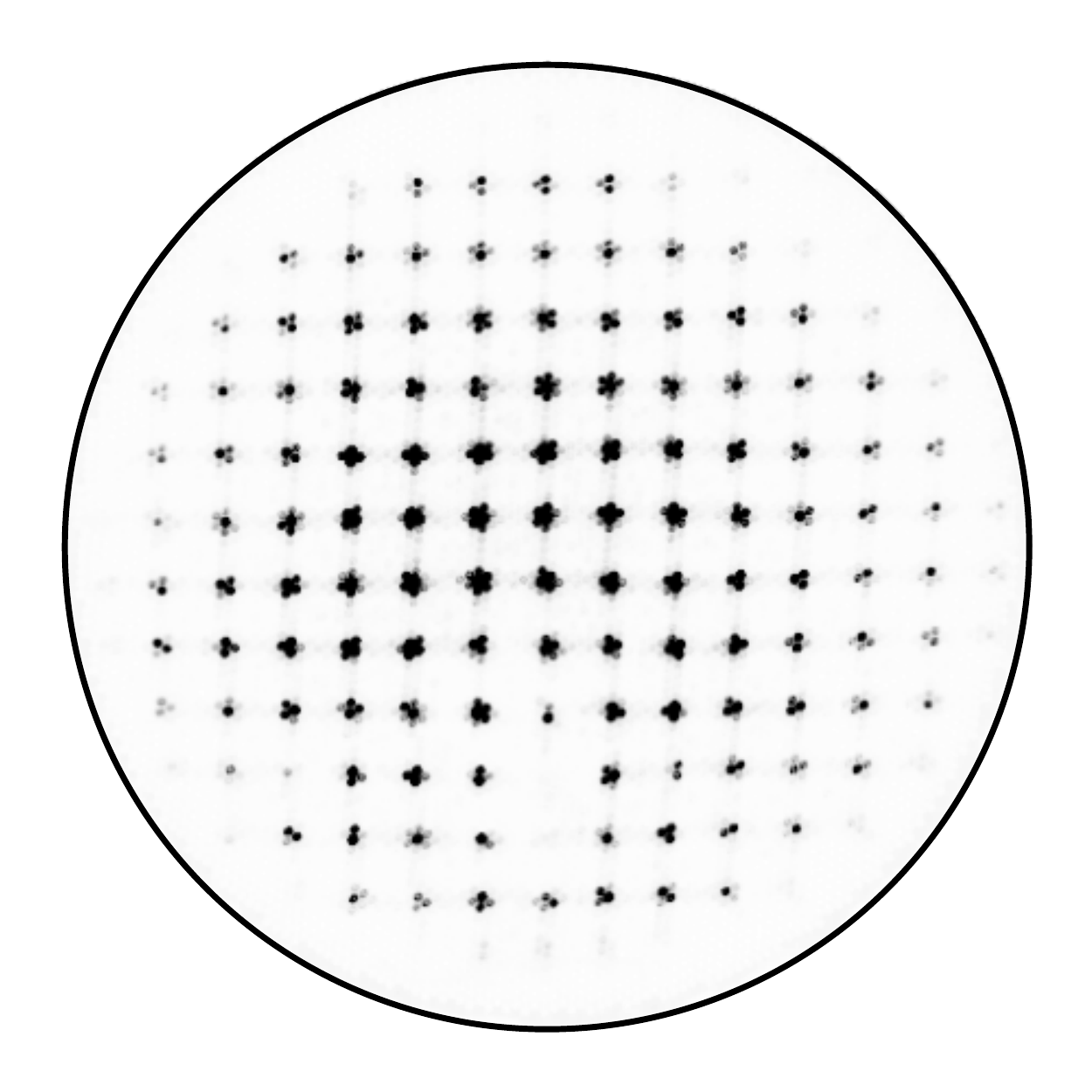}\includegraphics[width=0.49\linewidth]{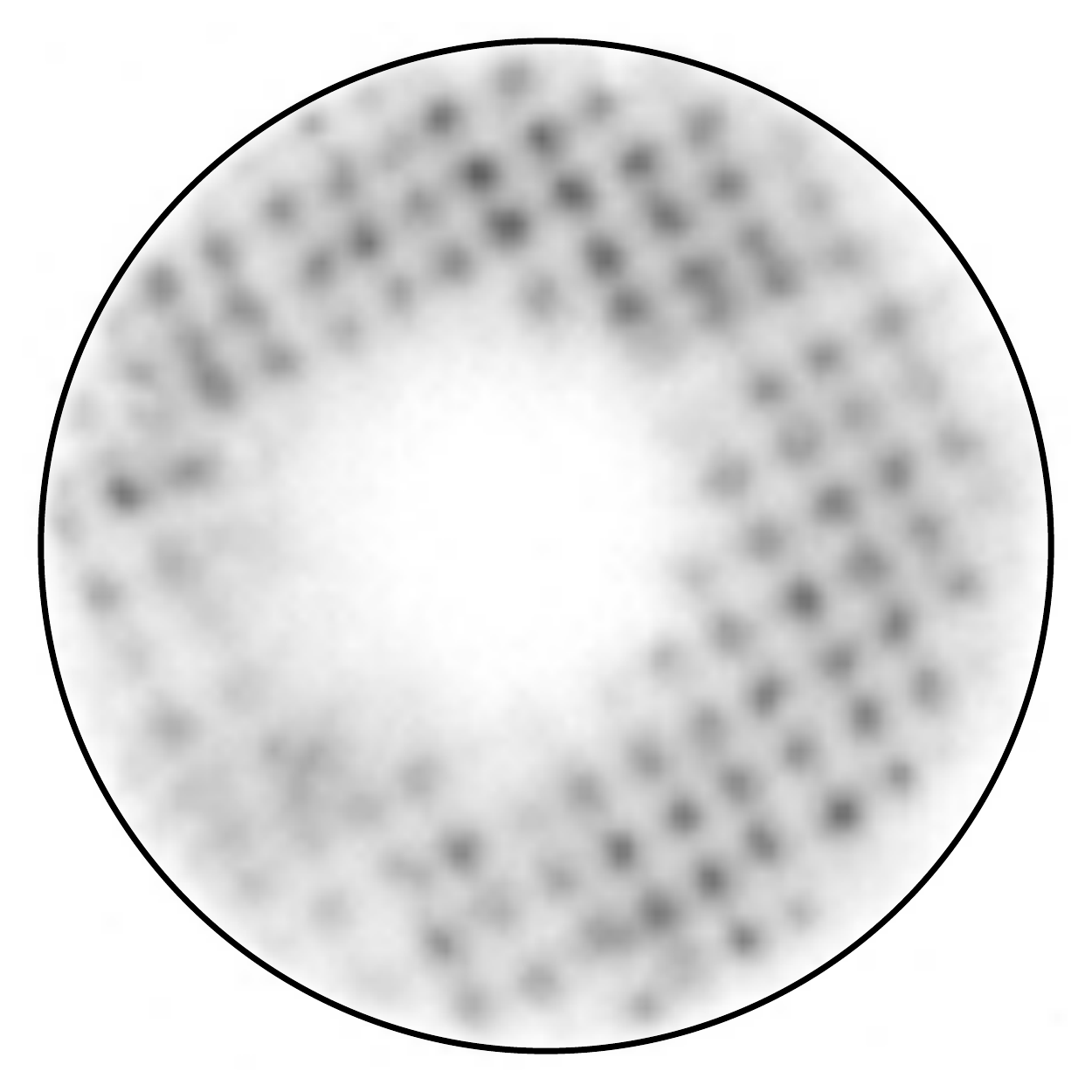}}
\caption{(top) Ray trace of miniature Shack-Hartmann sensor, where A is the starlight from the telescope (f/8 at AAT cassegrain focus), B is a $3$\,mm diameter collimating lens, C is a $110\,{\rm \mu m}$ pitch mircolens array, and D is the polished front face of a $1.5$\,mm polymer imaging bundle. (left) Image of a prototype miniature Shack-Hartman sensor illuminated with a $630$\,nm laser at f/8 (top). Dust inside the wavefront sensor blocks $\sim2$ of the lenslets. (right) On-sky demonstration of the same wavefront sensor as shown in (left) translating the spot pattern from the focal plane to a high frame rate camera (sCMOS). This stacked image is equivalent to a $1$\,s exposure. The de-centring of the pupil is due to a small tip/tilt in the optical alignment of the wavefront sensor to the focal plane. The wavefront sensor's field-of-view (sub-aperture spacing) is $\sim2$~arcsec. Both images have been inverted in colour, with a circle drawn to show the boundary of the imaging bundle.}
\label{fig:PIB_wfs}
\end{minipage}
\end{figure}

The use of polymer imaging bundles becomes an attractive solution to distributed wavefront sensors, where each bundle has miniature Shack-Hartmann fore-optics, relaying the wavefront information to a high frame rate camera situated off the focal plane. These miniature Shack-Hartmann wavefront sensors \citep[mini-SHWFS;][]{2014SPIE.9151E..4TG} were demonstrated on the $3.9$\,m AAT in December 2015 (see Figure \ref{fig:PIB_wfs}; a full description of these devices is provided by Goodwin et al. \emph{in preparation}). 

The $1.5$\,mm polymer imaging bundle, with its $16\,{\rm \mu m}$ cores and high filling factor, is able to critically sample the Shack-Hartmann spots of the wavefront sensor. A laboratory closed-loop adaptive optics experiment was carried out to quantify the suitability of the polymer imaging bundle in wavefront sensing (see Figure \ref{fig:PIB_wfs}). The set-up for the experiment was as follows: (1) Collimate the beam of a $635$\,nm laser to a beam width of $12.5$\,mm. (2) send the collimated laser beam through a rotatable phase-screen with effective seeing of $0.78$~arcsec. (3) image the pupil of the collimated beam onto an ALPAO 97-15 Deformable Mirror (DM). (4) re-image the pupil plane through a $100\,{\rm \mu m}$-pitch microlens array, resulting in the $1.5$\,mm polymer imaging bundle ($1$\,m length) receiving a $14\times14$ Shack-Hartmann spot pattern, which contains the wavefront information at the pupil plane. (5) re-image the end-face of the $1.5$\,mm polymer imaging bundle with $2\times$ magnification onto an Andor Zylar 5.5 MP sCMOS detector. (6) Control the rotation speed of the phase-screen, the SCMOS detector, and the DM via "S/W ALPAO Core Engine (ACE)" software using {\sc MATLAB/C++}. The software was customised to fit this experiment, which was able to run at $\sim250$~frames~per~second in closed-loop mode. The latency in the experiment was found to typically be $2$-$3$ frames. Optically, the set-up was designed to work in $f/10$, and to mimic the University of Hawaii's $2.2$\,m telescope. 

From Figure \ref{fig:PIB_wfs} (bottom) it is possible to see that the $1.5$\,mm polymer imaging bundle can indeed critically sample the Shack-Hartmann spot pattern as part of a closed-loop adaptive optics instrument, enabling the the Shack-Hartmann spot pattern to be relayed to a nearby, yet arbitrarily located, sensor. Only at wind-speeds greater than $10$~ms$^{-1}$ does the wavefront improvement become negligible, with the RMS approaching that of an uncorrected wavefront (open-loop). Custom designs of polymer imaging bundles can be manufactured depending on system requirements, however this is somewhat limited without the ability to self-manufacture as access to propriety polymers and equipment is restricted.

In addition to being used for high-order wavefront sensing, the polymer imaging bundles are also suited to distributed low-order wavefront sensing for use in ground-layer AO \citep[GLAO;][]{2010Natur.466..727H} or the monitoring of natural guide stars in laser-tomography AO \citep[LTAO;][]{2002ApOpt..41...11V}, which would also benefit any attempt of laser-tomography MOAO with either Rayleigh or Sodium laser guide stars.

An alternative method in wavefront sensing is the curvature sensor \citep{1988ApOpt..27.1223R,1993JOSAA..10.2277R}, where the intensity image of the intra- and extra-focus are simultaneously obtained, allowing the second derivative of the wavefront to be measured. The Visible and Infrared Survey Telescope for Astronomy (VISTA) built into its design the ability to obtain the intra-focus of one star and the extra-focus of another (but nearby) star with two respectively independent CCDs \citep{2003SPIE.4842..231P}. The assumption is that differential aberrations between the two stars are negligible. However, most AO systems that are based on curvature wavefront sensors \citep{1998PASP.110.152R,2003SPIE.4839..174A,2008SPIE.7015E..64W} use a resonant vibrating mirror with sinusoidal piston variations to obtain the intra- and extra-focuses. This is a very delicate procedure, and limits the adaptability to MOAO. Based off the design proposed by \citet{2010PASP..122...49G}, another method is to use two imaging bundles for each curvature sensor, with a beam splitter projecting the intra- and extra-focus images of a star onto each bundle. The imaging bundles then feed into a sensitive high frame-rate camera off the focal plane, and the intensity images are analysed to obtain the wavefront information. A full description of these devices and recent on-sky tests are given by \citet[][\emph{in preparation}]{2014SPIE.9148E..2EZ}. 


\begin{figure*}
\centering
\includegraphics[width=0.425\linewidth]{./_PIB_miniSHWFS_slopes.pdf} \hspace{1cm}
\includegraphics[width=0.425\linewidth]{./_PIB_miniSHWFS_flux.pdf} \\
\vspace{0.4cm}
\includegraphics[width=0.8\linewidth]{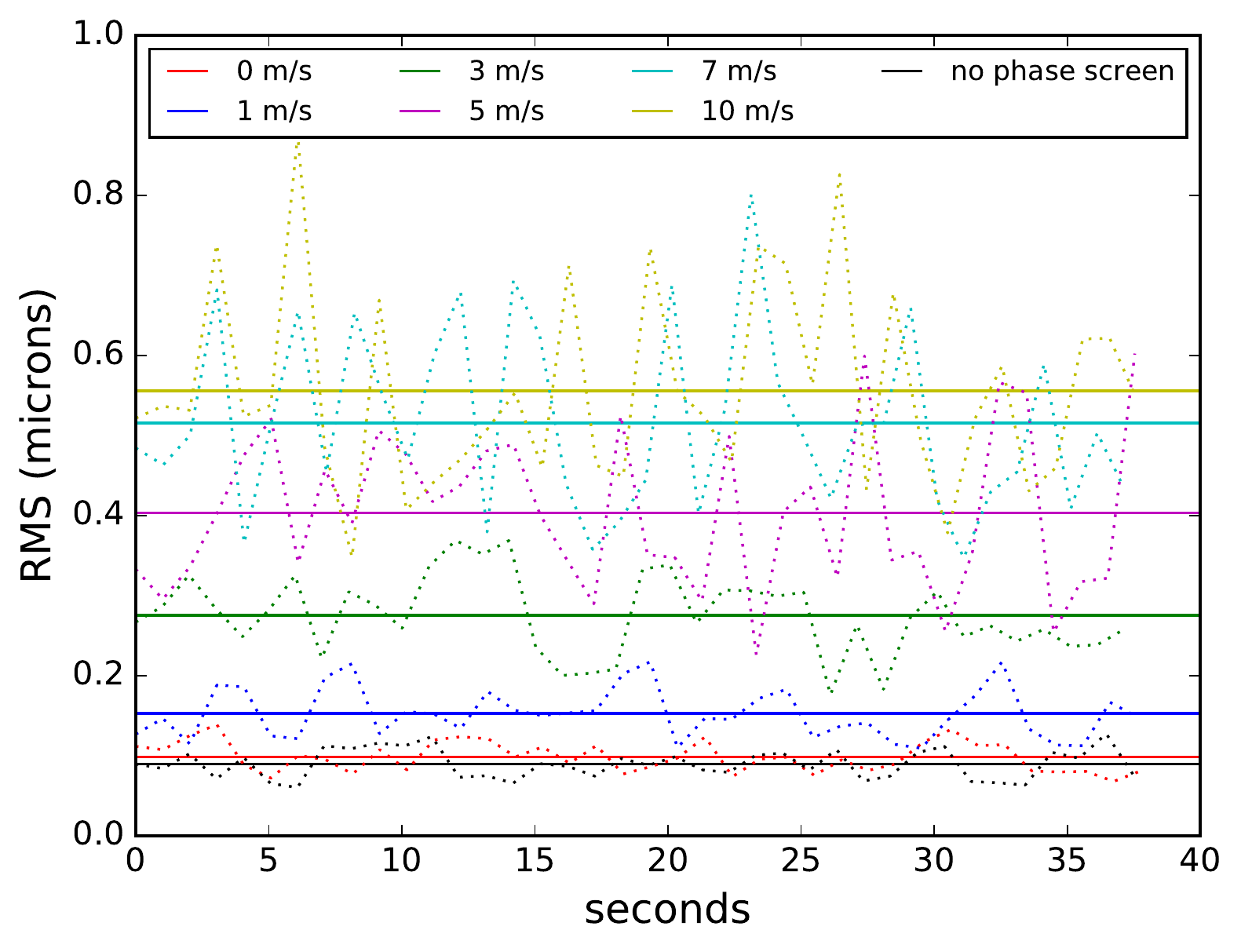}
\caption{Laboratory closed-loop adaptive optics data using a $1.5$\,mm polymer imaging bundle as part of a mini-Shack-Hartmann wavefront sensor (mini-SHWFS, see text for setup description). (top left) The $14\times14$ Shack-Hartmann spot pattern relayed by the $1.5$\,mm polymer imaging bundle. The centre-of-mass of each spot is equated in relation to the centre of each zonal box, with the offsets represented by red arrows. A mask is applied to skip zones/spots that have too little flux. Dust inside the mini-Shack-Hartmann sensor results in some zones to the left of the bundle being skipped. (top right) The flux map showing the masked out zones as black zones. (bottom) Closed-loop results for six different simulated wind speeds of the $0.78$~arcsec phase screen, in addition to no phase screen. The RMS is a measurement of the error compared to that of a flat wavefront (best image quality). The small RMS difference between the $0$~ms$^{-1}$ and no phase screen is a verification that the mini-SHWFS is successful. There is little to no wavefront improvement for wind speeds above $10$~ms$^{-1}$, with the RMS increasing to the uncorrected wavefront due to system latency ($2$-$3$ frames).}
\label{fig:PIB_AO}
\end{figure*}

\vspace{0.8cm}
\subsection{Other applications}

In addition to being used for field acquisition, guiding and wavefront sensing, polymer imaging bundles are also suited to a variety of other astronomical applications that require the efficient and flexible translation of images over short distances. One such application is in multi-object narrowband imaging, where each imaging bundle is positioned over a target and has a miniature tuneable narrowband filter that is tuned for its target (e.g. a galaxy's redshift). The ability to obtain high-spatial resolution narrowband images (e.g. \Ha) of $10^5$ independent galaxies over a survey lifetime can provide a dataset that is unparalleled in understanding the star formation morphology of a galaxy as a function of its environment. An issue if using the polymer bundle presented here is the spectral absorption features of the PMMA. There is a large absorption feature at $\sim740$~nm (see Figure \ref{fig:PIB_transmission}), which would be detrimental to \Ha\ observations of galaxies at redshifts, $0.06\lesssim z \lesssim 0.2$. Depending on the science in question, different polymers can be used to minimise spectral features, whilst still retaining other benefits such as high fill-factors. \citet{2013ISRN.785162..22A} provides a review of the common polymers used in the manufacture of fibre optics, with CYTOP having the best spectral transmission (high and flat; Figure \ref{fig:PIB_transmission}) in addition to operating at near-infrared wavelengths.

It is also possible to envisage an instrument that has distributed miniature aperture masks \citep{2000PASP..112..555T} over the focal plane, with the intensity of each interference pattern projected onto the polymer imaging bundles, all feeding a sensitive high frame-rate camera off the focal plane. This application would work much better if the polymer used in the bundles is one that is efficient at wavelengths in the near/mid-infrared, such that extrasolar planet detection by imaging is feasible \citep{2005Natur.433..261B,2012SPIE.8445E..06I}. Multi-object speckle imaging \citep{2002RvMP...74..551S} is also feasible with distributed polymer imaging bundles without fore-optics, again with all bundles feeding a sensitive high frame-rate camera off the focal plane. This would be useful in a fast survey to understand the fraction of multiple star systems \citep{2013ARA&A..51..269D}.

Over time, an increasing number of telescope instruments are achieving multi-object functionality as they aim to maximise the available science over the focal plane, so it is only natural to keep thinking of feasible ways to make use of the power and large focal plane of the ELTs to come. There are likely many more applications of polymer imaging bundles for astronomical applications, and we leave the reader to imagine. 


\vspace{0.8cm}
\section{Conclusions} \label{sec:conc}

Imaging bundles are a useful way of translating spatially coherent images, and over short distances, imaging bundles made from polymer can be more efficient and practical than than those made from silica. We performed laboratory characterisation of ESKA PMMA imaging bundles, and explored their various applications in the context of astronomical instrumentation. These cheap polymer imaging bundles ($\sim\$10$~per~metre) remain flexible as a single $1.5$\,mm diameter unit with $7095$ cores providing a spatial filling factor of $92$\%. Their R-band transmission was measured to be $\sim85$\%~per~metre, and their flat-field response yields a radial decrease of $\sim3$\% towards the outer cores. With a maximal NA of $0.5$, the polymer imaging bundles are well suited to a range of telescope focal ratios. As most applications will image the end-face of the imaging bundles with a camera, any FRD can be accounted for in the camera design. 

The polymer imaging bundles characterised here are already in use for field acquisition and guiding as part of the astronomical instrument, SAMI, and in the near-future, TAIPAN. This application is the most obvious use of spatially coherent imaging bundles, though other demonstrated applications include multi-object miniature wavefront sensing. Polymer imaging bundles are highly customisable in design (including which polymers to use depending on desired wavelength ranges) making them well suited to a range of astronomical applications, such as multi-object narrow-band imaging, multi-object aperture masking and multi-object speckle imaging. There are of course many other applications not yet explored, and the use of polymer imaging bundles in future astronomical instruments has potential to become routine.


\vspace{0.4cm}
\section{Acknowledgments}

We would like to thank Chris Betters, Simon Gross, Alexander Arriola, David Brown, and the staff at the Anglo-Australian Telescope for their help in the characterisation and commissioning of the different instruments that make use of the polymer imaging bundles. Most of the characterisation work was performed at the Sydney Astrophotonic Instrumentation Laboratory, funded by an ARC Laureate Fellowship awarded to Joss Bland-Hawthorn.

This research was also conducted in partnership with the Australian Research Council Centre of Excellence for All-sky Astrophysics (CAASTRO), funded under the Australian Research Council (ARC) Centre of Excellence program (project number CE110001020), with additional funding from the seven participating universities and from the NSW State Government's Science Leveraging Fund.


\vspace{0.4cm}

\bibliography{./_papers.bib}



\end{document}